\documentclass[epj]{svjour}
\usepackage[pdftex]{graphics}
\usepackage{color}
\usepackage{amsmath}
\usepackage{amssymb}
\usepackage{amsfonts}
\begin{document}
\title{History-dependent growth and reduction of the ripples formed on a swept granular track}
\author{Shunto Hata\inst{1}\thanks{\emph{Present address: Division of Earth and Planetary Sciences, Kyoto University, Kyoto 606-8502, Japan} } \and Makoto Katsura\inst{1} \and Hiroaki Katsuragi\inst{1}
%
}                     
%
%
\institute{Department of Earth and Space Science, Osaka University, 1-1 Machikaneyama, Toyonaka 560-0043, Japan}
\date{Received: date / Revised version: date}
%
\abstract{
When a solid object or wheel is repeatedly dragged on a dry sandy surface, ripple patterns are formed. Although the conditions to form ripple patterns have been studied well, methods to eliminate the developed ripple patterns have not been understood thus far. Therefore, history-dependent stability of the ripple patterns formed on a sandy surface is investigated in this study. First, the ripple patterns are formed by sweeping the flat sandy surface with a flexible plow at a constant speed. Then, the sweeping speed is reduced, and the variation of ripple patterns is measured.  As a result, we find that the ripple patterns show hysteresis. Specifically, the increase in amplitude of ripples is observed when the reduced velocity is close to the initial velocity forming the ripple pattern. In addition, splitting of ripples is found when the reduced velocity is further decreased. From a simple analysis of the plow's motion, we discuss the physical mechanism of the ripple splitting.
\PACS{
      {PACS-key}{discribing text of that key}   \and
      {PACS-key}{discribing text of that key}
     } 
} 
\maketitle
\section{Introduction}

When a wheeled vehicles repeatedly passes over an unpaved sandy road, corrugated ripple patterns are gradually developed on the road. This phenomenon is called washboard road development and often observed in dry conditions. Although the horizontally flat road is gravitationally stable, the flat surface becomes unstable by the repeated sweeping. 
Once the ripple patterns are developed on the road, it significantly reduces the comfortability of the driving. Moreover, it could also increase the possibility of traffic accidents. Therefore, undestanding and controlling the ripple development on a sandy surface have been challenging problems for many years~\cite{K1963,J1973,Fay:2001,S2006,daSilva:2019,C2020}. Besides, there are many kinds of ripple patterns around us that are generated by more or less similar mechanism. Examples include ripples mark in the desert~\cite{R1941}, mogul formation on slopes~\cite{J2002}, and rail corrugation~\cite{Y2002,P2011,G2008}. In particular, the rail corrugation has been extensively studied in the engineering field because it has a significant impact on our daily lives and economy.

Ripple formation process can be regarded as an instability of a flat surface. Thus, for example, this phenomenon has been studied based on the modeling in terms of nonlinear dynamics~\cite{J2001,D2000}. The first paper discussing washboard road was written by Mather~\cite{K1963}. At that time, it was considered that bouncing was necessary to form ripples ~\cite{K1963,G1989}. Therefore, most of the researchers considered that the suspension of cars passing over the road played a crucial role to develop the ripples. Later, some recent studies revealed that ripples could be generated by simply sweeping a granular surface with a plow that always contacting with the granular surface~\cite{N2007,A2009}. Furthermore, a recent study reports that even a single sweeping on a granular or viscoplastic layer can trigger the corrugation formation~\cite{Hewitt:2012}. Recently, the plows have been frequently used to investigate ripple-formation process because plowing is simpler than wheel rolling. For instance, a model for ripple growth was proposed by analyzing the forces acting on the plow from the granular surface~\cite{B2011,B2013}. One of the most important features of the ripple-formation process is the critical velocity below which ripple patterns cannot be developed~\cite{A2009}. The critical velocity can be characterized by a dimensionless number similar to the Froude number. By these previous studies, our understanding for the onset criterion and formation mechanics of the ripple patterns has been developed. 

However, these previous studies have not focussed on the obliteration of the ripple patterns. Because the flat surface is preferred for usual roads, we should understand the way to reduce the ripple development. Obviously, the simplest way to prevent the ripple formation is slowing down the sweeping velocity below the critical value. However, the slow sweeping takes a long time. If we can effectively obliterate the ripples by slightly varying the sweeping velocity, such a procedure would be a useful method to maintain smooth flat surfaces. Because both of the ripple formation and reduction processes depend on the sweeping velocity, its history-dependent ripple deformation caused by the velocity variation is a key issue. To find an efficient way to reduce the ripple patterns, we have to understand the history dependence of the surface processes. Specifically, how slow we should reduce the sweeping velocity to eliminate the existing ripples should be examined. Therefore, in this study, we focus on the history-dependent stability of the ripples formed on a granular surface. First, the ripple patterns are formed on a flat granular surface by repeatedly sweeping it by a rotatable plow at a constant velocity. Then, the sweeping velocity is reduced and the deformation of the ripple patterns is observed. We measure the amplitude and wavenumber of the developed ripple patterns, and discuss a simple mechanics which can explain what is happening on the plowed surface.

\section{Experiment}
\subsection{Experimental apparatus}
\begin{figure}
  \includegraphics{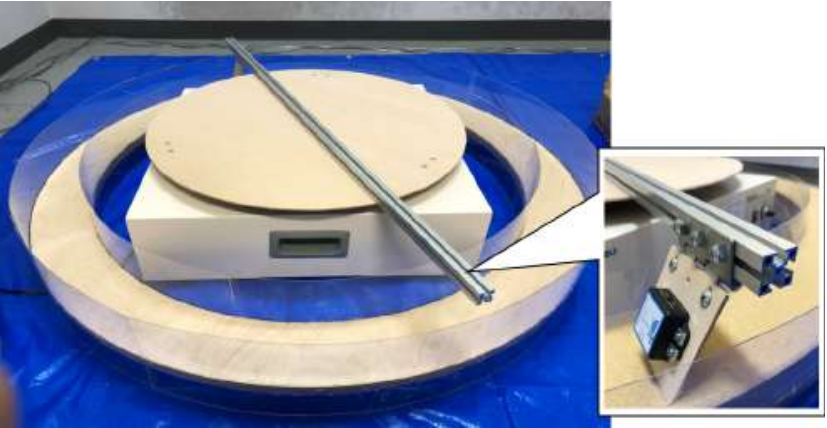}
  \caption{Photos of the experimental apparatus. The large photo shows the entire setup. The small insert shows the plow attached to the aluminum frame. Note that the plow itself can pivot because it is mounted on the bar by a hinge. The plow runs over the granular layer by rotating the frame.}
  \label{fig:setup}
\end{figure}

\begin{figure}
  \begin{center}
    \includegraphics{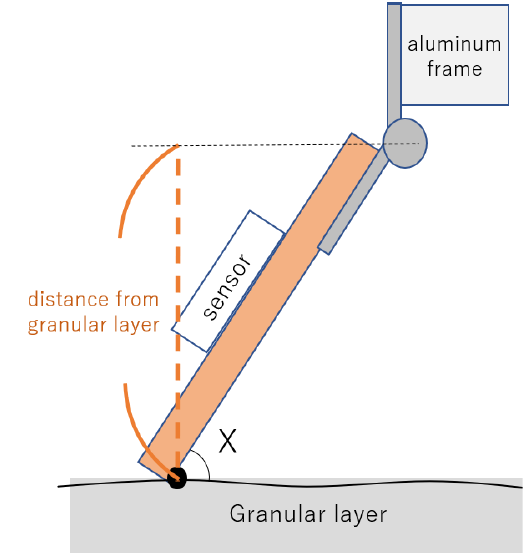}
  \end{center}
  \caption{Schematic view of the geometry around the plow and the sensor. We calculate the distance from the granular layer to the pivot by using the data measured by the sensor at 100~Hz data rate.}
  \label{fig:measurement}
\end{figure}
  
Figure~\ref{fig:setup} shows the experimental apparatus. It consists of a circular track and a rotating unit (Shimadzu-rika, MO-60). The track is filled with Toyoura sand (grain size: 100--300~$\mu$m, and true density: $2.6\times 10^3$~kg~m$^{-3}$), and the outer and inner diameters of the track are 1 and 0.8~m, respectively. Vertical thickness of the sand layer is approximately 40~mm. The axis of the rotating unit is placed at the center of the circular track. The aluminum frame (1~m in length) is mounted on the rotating unit. We use a wooden plow attached to one end of the frame to sweep the granular surface. The width, length, and thickness of the plow are 70~mm, $R=80$~mm, and 3~mm, respectively. The plow can freely rotate with respect to the frame by using a hinge between plow and frame; flexible plow. By rotating the frame, the plow runs over the granular surface with a constant speed ranging from 0.23 to 1.21~m~s$^{-1}$ (4.9--25.8~rpm). In this experiment, we use the wireless inclinometer sensor (Wit motion, BWT901CL) mounted on the plow which measures the angle $X$ and the position of the plow at $100$~Hz sampling rate. Total mass of the plow including the sensor is $M=40$~g. Figure~\ref{fig:measurement} shows the plow and the sensor configuration. The angle $X$ is a key parameter in this study because we can compute various quantities from $X(t)$ data; e.g., surface height of the granular layer can be computed as $h=R(\sin X_0 - \sin X)$ where $X_0\simeq 60^{\circ}$ corresponds to the reference height. Basically, the surface profiles are characterized by these data. Based on these experimentally obtained data, we discuss the dynamics of the plow which determines the granular surface profile.

\subsection{Experimental procedure}
First, we prepare the flat granular surface by sweeping its surface with the fixed (not flexible) plate. In this preparation stage, the lower-edge height of the fixed plate is kept constant. Thus, we can level the surface by dragging the fixed plate at a sufficiently low speed. 

After preparing the flat surface, we detach the fixed plate from the frame and let the flexible plow run over the surface at a constant velocity, which is defined as \textit{appear velocity}, $v_\mathrm{ap}$. 
After a sufficiently long-time sweeping (typically 15~min.), the ripple pattern is developed on the surface and approaches the steady state. Then, the amplitude and wavenumber are measured by the plow's motion. To compute the rotation rate, we measure the period necessary for ten rotations by using a stopwatch. Note that this stage still corresponds to the preparation process in this study. 

Finally, the plow is dragged at a \textit{reduced velocity} $v_\mathrm{red}$ which is less than $v_\mathrm{ap}$. After sufficient rotations, the surface pattern approaches the steady state. To characterize the ripples, we measure the amplitude and wavenumber of the steady surface profiles. Steadiness of the ripple patterns can be confirmed in the temporal development of ripple amplitude (see Appendix A).

\section{Result}
\subsection{Appearing ripples}

\begin{figure}
  \includegraphics{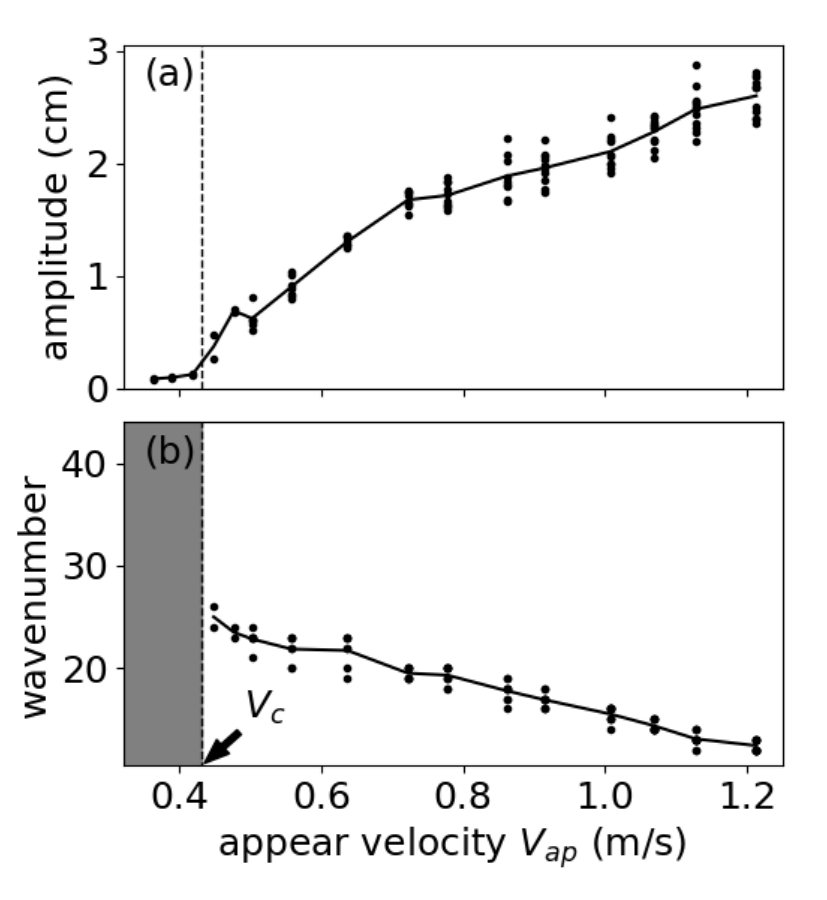}
  \caption{(a) Average amplitude and (b) wavenumber of the ripples developed on the granular surface with $v_\mathrm{ap}$. Each symbol indicates individual experimental result. Solid lines connect the average values of the same $v_\mathrm{ap}$ data. The region below the critical velocity $v_\mathrm{c} = 0.43$~m~s$^{-1}$ is shaded grey in (b).}
  \label{fig:resap}
\end{figure}

\begin{figure}
  \includegraphics{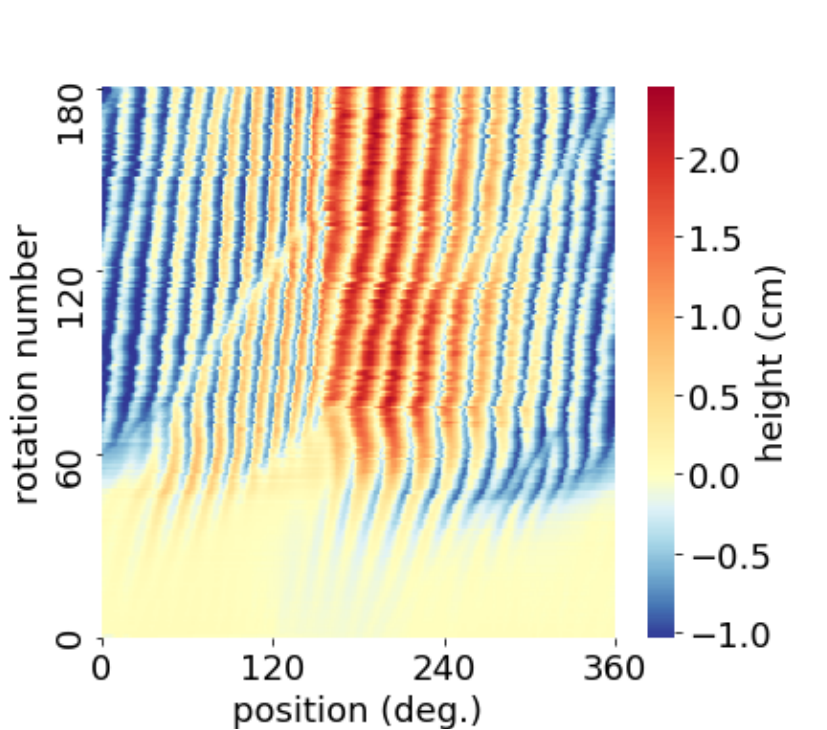}
  \caption{The space-time plot showing the growth of a ripple pattern produced from the horizontally flat surface by $v_\mathrm{ap}=0.78$~m~s$^{-1}$.}
  \label{fig:up}
\end{figure}

In Figure~\ref{fig:resap}, amplitude and wavenumber of the ripple patterns formed by $v_\mathrm{ap}$ are displayed. Figure~\ref{fig:resap}(a) illustrates that the amplitude is an increasing function of $v_\mathrm{ap}$. The critical velocity $v_\mathrm{c}$ below which ripples cannot be developed is about 0.43~m~s$^{-1}$. The value of $v_\mathrm{c}$ depends on the system parameters. For example, Bitbol et al. obtained a scaling relation suggesting that $v_\mathrm{c}$ is proportional to $(M/w)^{1/4}$, where $M$ and $w$ are the plow's mass and width, respectively~\cite{A2009}. Because we fix these parameters in this experiment, it is difficult to confirm this relation. Instead, we systematically vary the sweeping velocity and observe the associated (history-dependent) deformation of the ripple patterns, in this study. Figure~\ref{fig:resap}(b) shows the $v_\mathrm{ap}$ dependence of the wavenumber (number of peaks on the track). According to Figure~\ref{fig:resap}(b), the wavenumber has a negative correlation with $v_\mathrm{ap}$. In other words, fast sweeping results in the long wavelength. The region below the critical velocity $v_\mathrm{c}$ is shaded grey. These observed results are qualitatively consistent with the previous study~\cite{T2017}.

Figure~\ref{fig:up} shows a typical space-time plot of the ripple development. Horizontal axis corresponds to the position (phase) in the track, and the vertical axis corresponds to the time (in the unit of the number of rotation). Because the track is circular, position can be denoted by the angle ($360$~deg.~means one rotation). The color indicates the height of the surface as labeled by the color scale. The instantaneous height is computed from the inclinometer sensor data. As seen in Figure~\ref{fig:up}, the ripple pattern gradually grows from the bottom (initial state) to the top at which the steady state is achieved.

\subsection{Reducing ripples}
\begin{figure}[htbp]
  \includegraphics{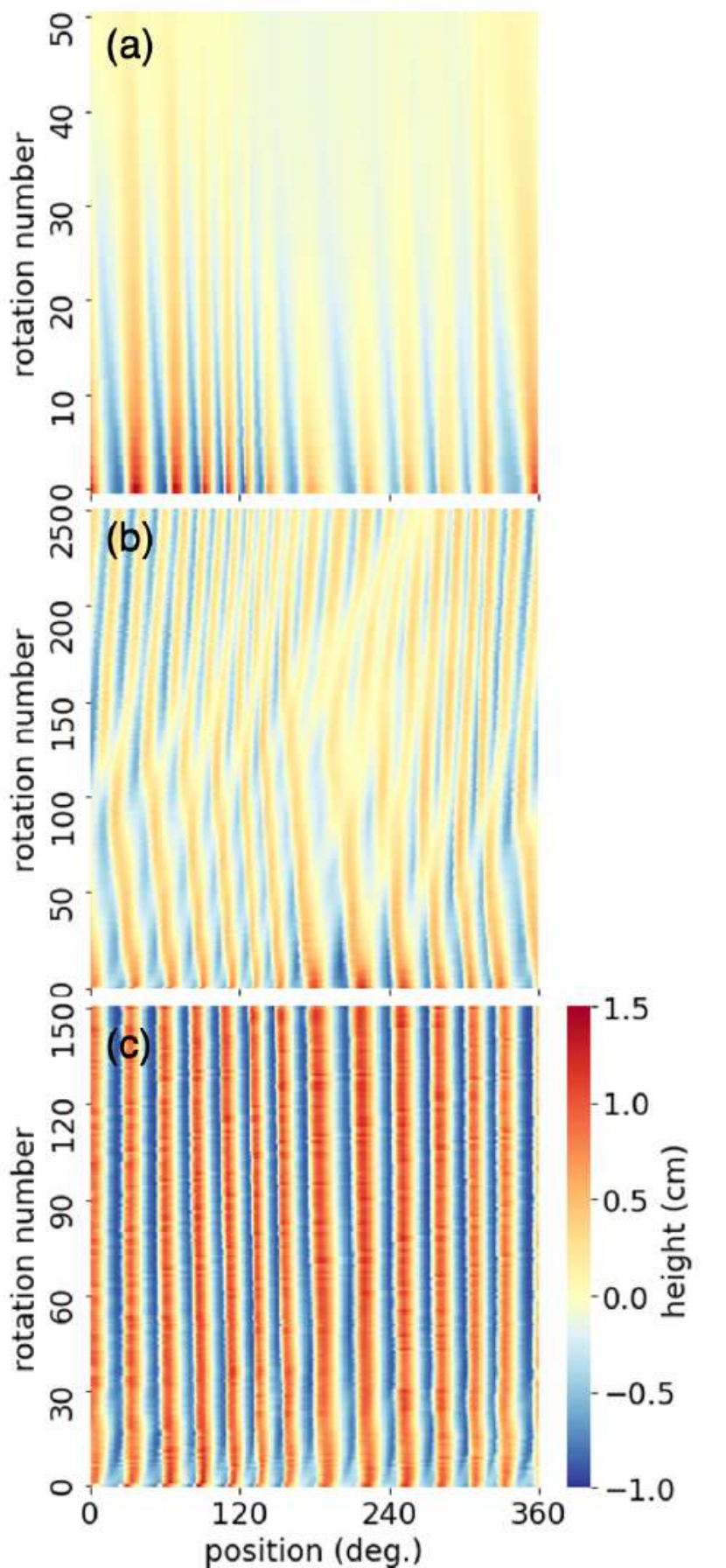}
  \caption{The space-time plots of three-type behaviors induced by the $v_\mathrm{red}$ plowing. The initial conditions of all these patterns are generated at the constant $v_\mathrm{ap}=1.21$~m~s$^{-1}$. The zero rotation number corresponds to this state. Then, the sweeping velocity is reduced to $v_\mathrm{red}$. Depending on $v_\mathrm{red}$, we observe three typical outcomes. (a) Disappear type: ripples vanishes at the slowest speed $v_\mathrm{red}=0.23$~m~s$^{-1}$. (b) Split type: the wavenumber increases at $v_\mathrm{red}=0.50$~m~s$^{-1}$. (c) Keep type: the wavenumber and amplitude do not vary at $v_\mathrm{red}=0.64$~m~s$^{-1}$.}
    \label{fig:disappear}
\end{figure}

Next, the sweeping velocity is reduced to $v_\mathrm{red}$ and the resultant ripple deformation is measured. As a result, we find that there are three types of ripple behaviors. Figure~\ref{fig:disappear} shows the space-time plots of various $v_\mathrm{red}$ cases (with the fixed $v_\mathrm{ap}=1.21$~m~s$^{-1}$). Figure~\ref{fig:disappear}(a) shows the slowest $v_\mathrm{red}$ case ($v_\mathrm{red}=0.23$~m~s$^{-1}$). In this case, the ripple pattern is obliterated. Therefore, this type of behavior is named \textit{disappear type}. On the other hand, in the fastest $v_\mathrm{red}(=0.64$~m~s$^{-1})$ case, the ripple characteristics do not vary (Figure~\ref{fig:disappear}(c)). This behavior is named \textit{keep type}. In the intermediate $v_\mathrm{red}(=0.50$~m~s$^{-1})$ case, the ripples split and namely the wavenumber of the ripples increases (Figure~\ref{fig:disappear}(b)). We name this behavior \textit{split type}.

\begin{figure}
  \includegraphics{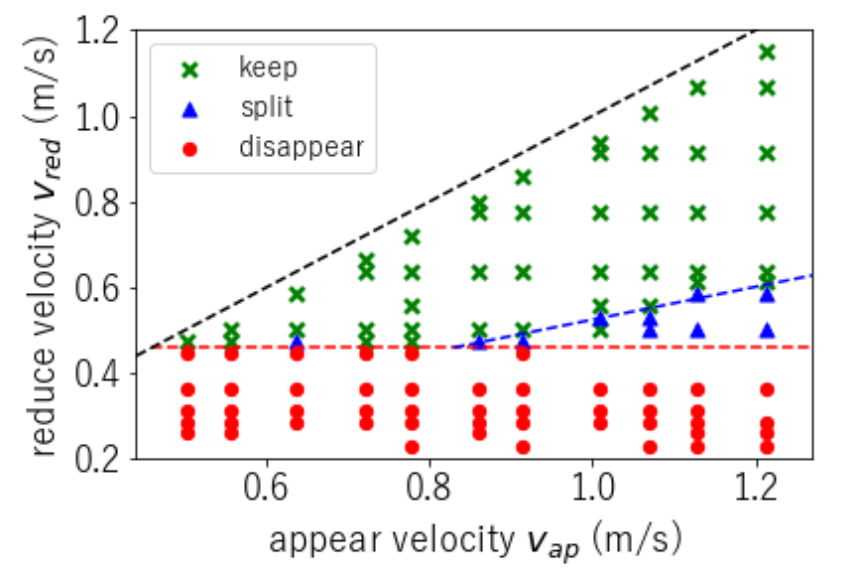}
  \caption{The phase diagram of the ripples variation with various combinations of $v_\mathrm{ap}$ and $v_\mathrm{red}$. The black dashed line represents $v_\mathrm{red}= v_\mathrm{ap}$. The red and the blue dashed lines are drawn to represent the boundaries of the phases.}
  \label{fig:souzu}
\end{figure}

We perform a set of experiments by systematically varying $v_\mathrm{ap}$ and $v_\mathrm{red}$ to make a phase diagram. Figure~\ref{fig:souzu} displays the obtained phase diagram. The horizontal and vertical axes indicate $v_\mathrm{ap}$ and $v_\mathrm{red}$, respectively. The black dashed line represents $v_\mathrm{red}=v_\mathrm{ap}$. Because the reduction of the ripples is aimed, all the experiments are carried out under the condition of $v_\mathrm{red} < v_\mathrm{ap}$. Thus, all the data plotted in the phase diagram are below the black dashed line. The disappear-type behavior is induced when $v_\mathrm{red}$ is below the red dashed line which corresponds to $v_\mathrm{c}=0.43$~m~s~$^{-1}$. Because this velocity level is independent of $v_\mathrm{ap}$, $v_\mathrm{c}$ is seemingly the universal velocity to completely eliminate the ripples. On the other hand, the keep-type behavior appears when $v_\mathrm{red}$ is relatively fast (close to $v_\mathrm{ap}$). Regarding the split-type behavior, we can observe this type of behavior between keep and disappear types in the relatively fast $v_\mathrm{ap}$ regime. The blue dashed line in Figure~\ref{fig:souzu} (phase boundary) is drawn just for the guide to the eye.

\section{Analysis and discussion}
\subsection{Amplitude analysis}
\begin{figure}
  \includegraphics{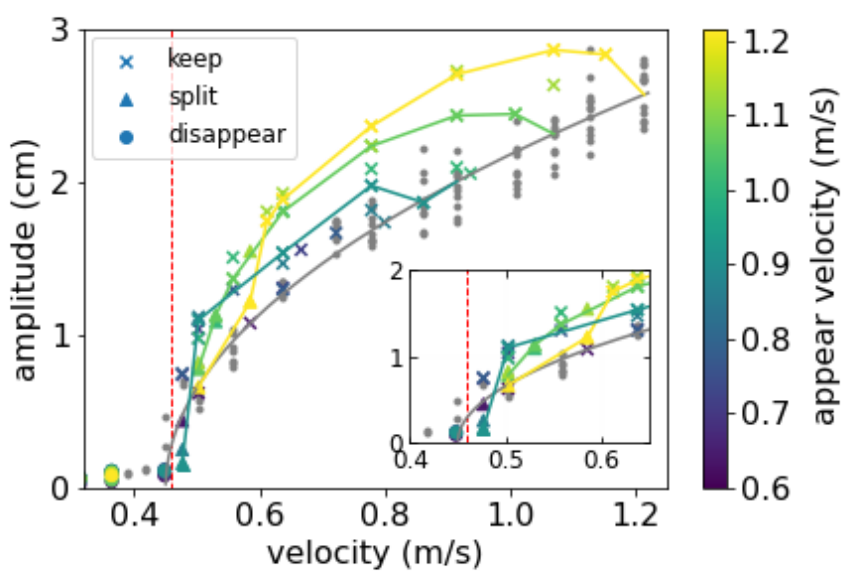}
  \caption{Amplitude variations by varying velocities. Grey plots and curve represent the appearing amplitude $A_\mathrm{ap}$ produced by $v_\mathrm{ap}$ from the flat surface. The horizontal axis corresponds to $v_\mathrm{ap}$ for gray data. Color plots represent the amplitude $A_\mathrm{red}$ produced by $v_\mathrm{red}$. The same color data indicate the same $v_\mathrm{app}$ but different $v_\mathrm{red}$ data. That is, the color lines connect the same $v_\mathrm{ap}$ data (yellow: 1.21~m~s$^{-1}$, green: 1.07~m~s$^{-1}$, blue: 0.86~m~s$^{-1}$). For the color data, the horizontal axis indicates $v_\mathrm{red}$. The inset shows the magnified plot.} 
  \label{fig:amphys}
\end{figure}

Figure~\ref{fig:amphys} shows the relation between appearing amplitude $A_\mathrm{ap}$ and reduced amplitude $A_\mathrm{red}$ in each steady state. The grey points represent the appearing amplitude $A_\mathrm{ap}$ produced by $v_\mathrm{ap}$. The colored curves represent the amplitudes $A_\mathrm{red}$  produced by the identical $v_\mathrm{ap}$ but various $v_\mathrm{red}$. The color indicates the magnitude of $v_\mathrm{ap}$. Namely, we connect the data of same $v_\mathrm{ap}$ ($v_\mathrm{ap}=1.21, 1.07,$ or $0.86$~m~s$^{-1}$) by the identical color curves. For example, the points connected by the yellow curve are all produced by $v_\mathrm{ap}=1.21$~m~s$^{-1}$ and various $v_\mathrm{red}$. Horizontal axis of Figure~\ref{fig:amphys} indicates $v_\mathrm{ap}$ and $v_\mathrm{red}$ for gray and color data, respectively. In the relatively fast regime ($v_\mathrm{red} \gtrsim 0.6$~m~s$^{-1}$), the reduced amplitude $A_\mathrm{red}$ is greater than the appearing amplitude $A_\mathrm{ap}$. This counterintuitive behavior suggests the complex history dependence (hysteresis) of the ripple deformation process. To the best of our knowledge, this is the first report of such a history dependence of the ripple deformation. 

In Figure~\ref{fig:amphys}, we find abrupt variation of the amplitude at the vicinity of $v_\mathrm{c}$. The inset of Figure~\ref{fig:amphys} displays the magnified plot illustrating the sudden jump of the amplitude. This peculiar behavior of the amplitude implies that insufficient reduction of the sweeping velocity is not a good idea to eliminate the already existing ripples. To completely erase the ripples, $v_\mathrm{red}$ should be reduced to the level of $v_\mathrm{c}$. If we slightly reduce the sweeping velocity, it might result in an adverse effect. In this sense, it is difficult to effectively eliminate the ripples even by using the history-dependent amplitude behavior. This result is practically unfortunate. However, when the splitting is induced, the amplitude significantly drops. Therefore, the physical mechanism governing the transition from keep-type to split-type should be further analyzed. 

\begin{figure}
  \includegraphics{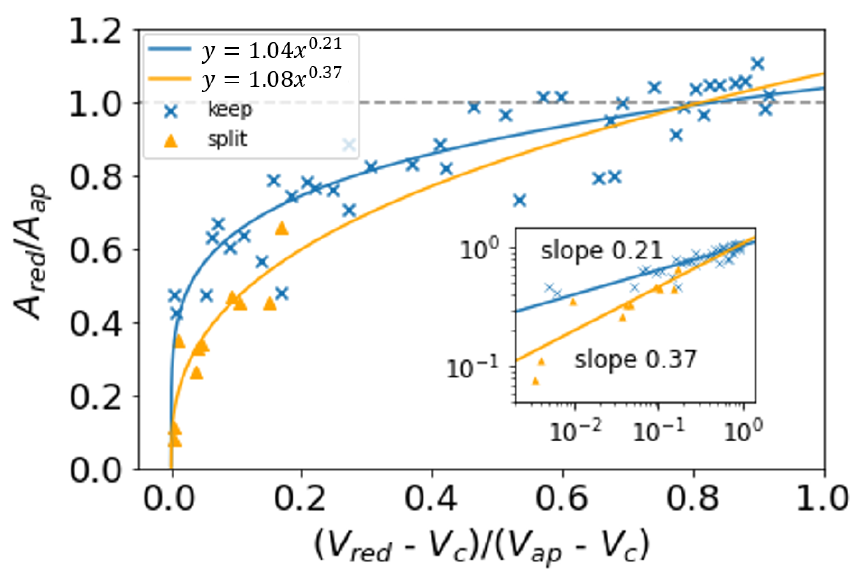}
  \caption{Amplitude ratio versus normalized velocity. The different behaviors can be confirmed between the keep type (cross marks) and split type (triangle marks). Solid blue and yellow curves are the fitting to keep and split types, respectively. The inset shows the double logarithmic plot of the same data.}
  \label{fig:ampnorm}
\end{figure}

To clearly see the difference between keep and split types, the amplitude and velocity data are normalized as $A_\mathrm{red}/A_\mathrm{ap}$ and $(v_\mathrm{red}-v_\mathrm{c})/(v_\mathrm{ap}-v_\mathrm{c})$, respectively. The relation between the normalized amplitude and velocity ratio is shown in Figure~\ref{fig:ampnorm}. The solid curves are the power-law fitting to keep-type and split-type data. The inset of Figure~\ref{fig:ampnorm} shows the double logarithmic plot of the same data. As shown in Figure~\ref{fig:ampnorm}, different scaling laws should be considered for keep and split types. If the granular bed is incompressible, the peak amplitude should be reduced by the splitting. Thus, the amplitude of the ripples can be efficiently reduced by the splitting process. When $v_\mathrm{red}$ is close to $v_\mathrm{ap}$ (and $v_\mathrm{red} \lesssim v_\mathrm{ap}$), however, eroding ability of the plow might be slightly weakened because the momentum of plow is slightly reduced as $Mv_\mathrm{red} \lesssim Mv_\mathrm{ap}$. This effect might enhance the amplitude of already existing ripple patterns. Therefore, the marginal slow down rather amplifies the amplitude as shown in Figure~\ref{fig:amphys}. If $v_\mathrm{red}$ is sufficiently small, wave splitting occurs. Although we do not completely understand this nonlinear behavior, the analysis of plow's rotation provides some clues to consider the physical mechanism. In the following, we discuss the rotational motion of the plow at the boundary between keep-type and split-type behaviors.

\subsection{Boundary between keep and split types}
\begin{figure}
  \includegraphics{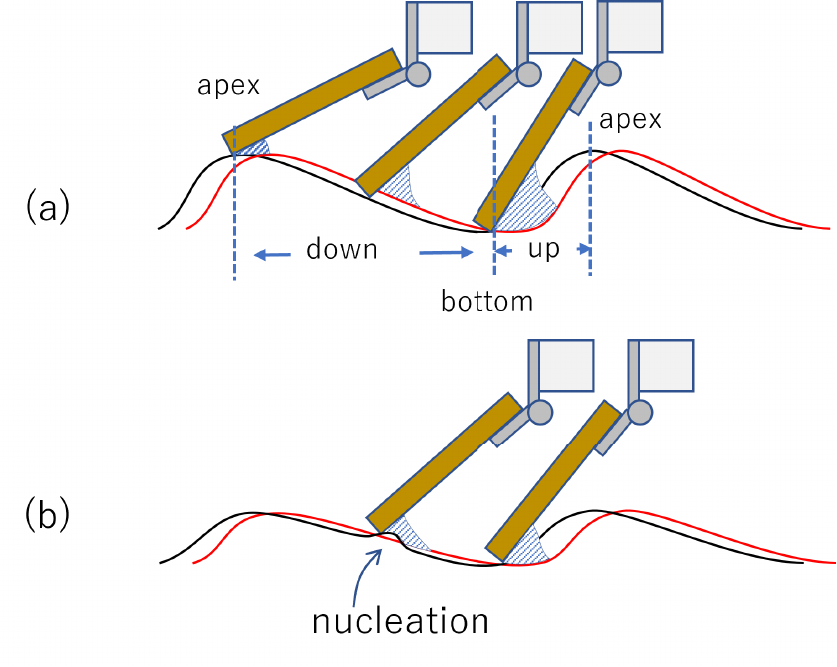}
  \caption{Schematic illustration of the sweeping plow on the granular surface: (a) keep-type ripples close to the splitting phase and (b)~nucleation resulting in the splitting. The plow moves from left to right. By the sweeping, the red profiles are deformed to the black profiles. That is, the surface profiles migrate to the opposite direction against the plow's motion. The shaded area shows eroded sand. (a) The decreasing- and increasing-height regimes are named down and up, respectively. (b) At the nucleation point, deposition of the eroded sand is triggered due to the small angular velocity (torque) of the plow.}
  \label{fig:profile}
\end{figure}

The onset condition of the splitting behavior is a key factor to understand the history dependence of ripple formation. 
Typical surface profile and sweeping plow in the keep-type phase close to the split-type phase is schematically shown in Figure~\ref{fig:profile}(a). In this phase, the asymmetric surface profile results in the backward migration as reported in~\cite{A2009}. The surface layer is eroded at the \textit{down} regime and the eroded grains are deposited at the \textit{up} regime (Figure~\ref{fig:profile}(a)).

By careful observation, we realized that most of the nucleations of the ripple splitting are triggered at the down slope as illustrated in Figure~\ref{fig:profile}(b). When the plow's torque becomes insufficient to erode the surface, deposition of the eroded sand begins instead of the eroding. To characterize the torque, the angular acceleration of the plow is evaluated. Particularly, we consider that the angular acceleration at the bottom (boundary of the down and up regimes (Figure~\ref{fig:profile}(a))) should be small when the eroding ability of the plow decreases. Therefore, in the following discussion, we focus on the average angular acceleration of the plow at the bottom, $\dot{\omega}_{\rm B}$, as an indicator of the eroding ability of the plow. 

\begin{figure}
  \includegraphics{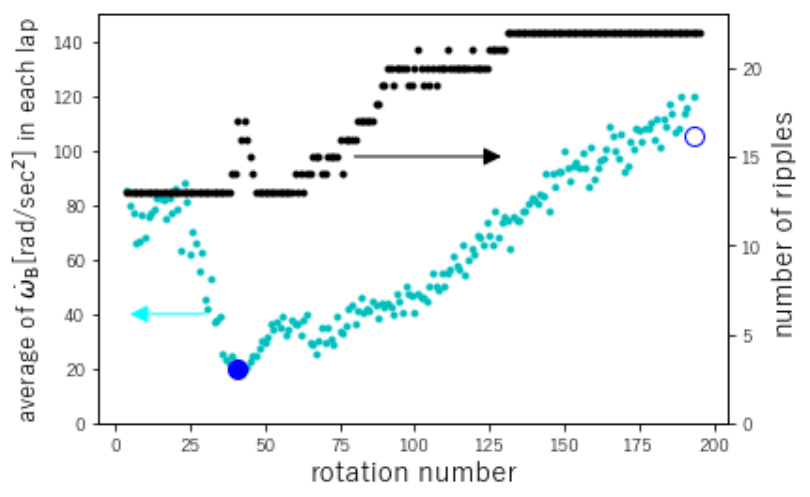}
  \caption{The average angular acceleration at the bottom, $\dot{\omega}_{\rm B}$, is plotted by green dots and the number of ripples is shown by black dots. A blue filled circle and a blue empty circle indicate the minimum and final values of $\dot{\omega}_{\rm B}$, respectively.}
  \label{fig:dot_omega_trend}
\end{figure}

Figure~\ref{fig:dot_omega_trend} shows the variation of $\dot{\omega}_{\rm B}$ (green dots) and the corresponding number of ripples in the track (black dots) at the splitting nucleation stage.The horizontal axis of Figure~\ref{fig:dot_omega_trend} denotes the rotation number after the sweeping velocity is set to $v_\mathrm{red}$. Note that the $\dot{\omega}_{\rm B}$ value shown in Figure~\ref{fig:dot_omega_trend} represents the average value per each rotation.  As seen in Figure~\ref{fig:dot_omega_trend}, the splitting (increase of the number of ripples) occurs approximately from 40 to 100 rotations. In this regime, $\dot{\omega}_{\rm B}$ is relatively small. We consider the low eroding ability characterized by small $\dot{\omega}_{\rm B}$ induces the wave splitting. After the splitting, $\dot{\omega}_{\rm B}$ gradually increases. Then, due to the relatively high eroding ability, the shape is stabilized. Therefore, we consider that the minimum and final values of $\dot{\omega}_{\rm B}$ can characterize the variation of the eroding ability and associated nucleation of the wave splitting. 

\begin{figure}
  \includegraphics{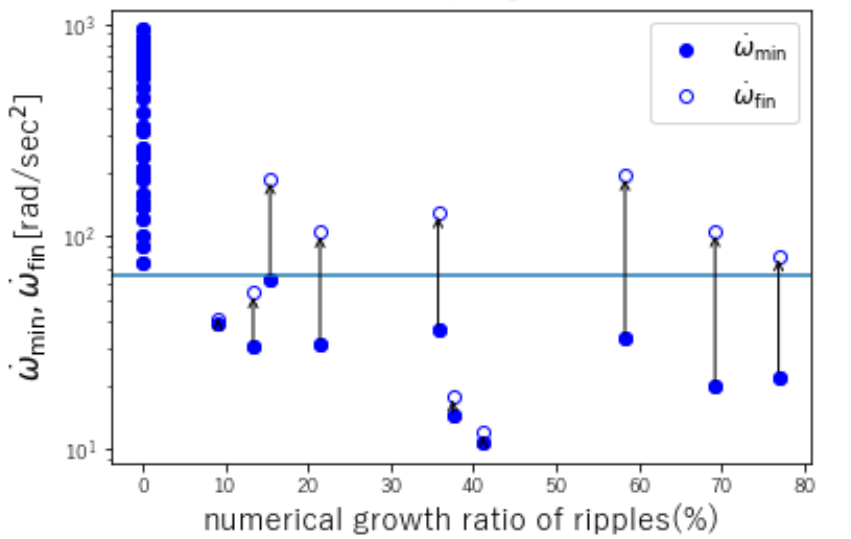}
  \caption{Filled circles show $\dot{\omega}_{\rm min}$ in both keep and split types. There seems a threshold value (65 rad~s$^{-2}$) as shown by the horizontal line. Empty circles show the final values $\dot{\omega}_{\rm fin}$ in the split type. Most of them exceed the threshold level.}
  \label{fig:dot_omega_scatter}
\end{figure}

Therefore, we define and measure $\dot{\omega}_{\rm min}$ and $\dot{\omega}_{\rm fin}$ as the minimum and final values of the $\dot{\omega}_{\rm B}$ as displayed by a blue filled circle and a blue empty circle in Figure~\ref{fig:dot_omega_trend}, respectively. In Figure~\ref{fig:dot_omega_scatter}, measured $\dot{\omega}_{\rm min}$ and $\dot{\omega}_{\rm fin}$ values are shown. The abscissa of Figure~\ref{fig:dot_omega_scatter} denotes the numerical growth ratio of ripples defined as, 
\begin{equation}
\mbox{Numerical growth ratio of ripples} 
= \frac{N_{\rm fin} -N_{\rm ini}}{N_{\rm ini}}, 
\end{equation}
where $N_{\rm ini}$ and $N_{\rm fin}$ are the initial and final number of ripples (during the sweep with $v_\mathrm{red}$), respectively. The data of $\dot{\omega}_{\rm min}$ for the keep type are aligned in the vertical axis because the number of ripples are kept constant (zero growth ratio) in the keep type. For simplicity, $\dot{\omega}_{\rm fin}$ of the keep type is omitted in Figure~\ref{fig:dot_omega_scatter}. 

According to Figure~\ref{fig:dot_omega_scatter}, when $\dot{\omega}_{\rm min}$ is smaller than a threshold value 65~rad~s$^{-2}$ (indicated by the horizontal line in Figure~\ref{fig:dot_omega_scatter}), splitting is induced (non-zero growth ratio). All the $\dot{\omega}_{\rm min}$ values of the keep type (zero growth ratio) are greater than this threshold value. Besides, most of the final values of the split type, $\dot{\omega}_{\rm fin}$, exceed the threshold value. This probably means the stabilization of ripples and suppression of further splitting. The minimum eroding ability $\dot{\omega}_{\rm min}$ can be an indicator for the nucleation. However, to evaluate the degree of instability by this indicator, we have to precisely measure the motion of plow. To find a more practical indicator, further analysis is necessary. Such a detailed analysis is a future problem.

\subsection{Discussion}

According to the obtained data and the analyzed result, the best way to reduce the ripple amplitude is reducing the sweeping velocity below $v_\mathrm{c}$. If $v_\mathrm{red}$ is close to $v_\mathrm{ap}$, the amplitude of ripple patterns could rather be enhanced. Based on Figure~\ref{fig:ampnorm}, we consider that $v_\mathrm{red}$ should satisfy $(v_\mathrm{red}-v_\mathrm{c})/(v_\mathrm{ap}-v_\mathrm{c}) \leq 0.2$ in order to efficiently reduce the amplitude. In this regime, splitting of ripples could be induced. 

The range of $v_\mathrm{ap}$ in this experiment is limited in $0.23$--$1.21$~m~s$^{-1}$ due to the technical limitations. Therefore, the range of splitting phase is very limited (Figure~\ref{fig:souzu}). To overcome the limitation, we have to build a larger experimental system. Such a large-scale experiment is possible future work.

\section{Conclusion}
In this study, we investigate the reduction process of granular ripple amplitude by decreasing the plow's sweeping velocity. As a result, we find three types of reducing processes: keep, split, and disappear. When $v_\mathrm{red}$ is close to $v_\mathrm{ap}$, the ripple patterns keep or enhance their shape. We call this regime keep type. In the split type, wavenumber of ripple patterns increases and amplitude decreases. The boundary between keep and split behaviors can be understood by considering torque acting on the plow. The new ripples appear when the total torque from sand becomes larger than that of gravity in down regime. This process relates to the definition of the onset of splitting. The complete disappear behavior arises when $v_\mathrm{red}$ is below the critical velocity $v_\mathrm{c}$. According to the experimental result, the critical velocity to develop ripple patterns on the flat surface is almost identical to the critical velocity below which the ripple patterns can completely be obliterated. That is, $v_c$ is the universal critical velocity. As mentioned above, the amplitude of the ripple pattern can be enhanced by slightly reducing the sweeping velocity. This means that the small reduction of the sweeping velocity is not an efficient way to reduce the ripple amplitude. To effectively reduce the ripple amplitude, $v_\mathrm{red}$ should be smaller than the value satisfying $(v_\mathrm{red}-v_\mathrm{c})/(v_\mathrm{ap}-v_\mathrm{c}) \leq 0.2$. 

It would be interesting to analyze the temporal variation of amplitude. In this study, we mainly analyze the steady state of the ripple patterns. Through the dynamical (time-dependent) analysis, we would be able to understand the details of the history-dependent ripple development.

\noindent 
\textbf{Funding information} The financial support of Japan Society for the Promotion of Science (JSPS) KAKENHI (18H03679 and 19H01951) is appreciated.

\section*{Appendix A: Steadiness of ripple pattern}
\begin{figure}
  \includegraphics{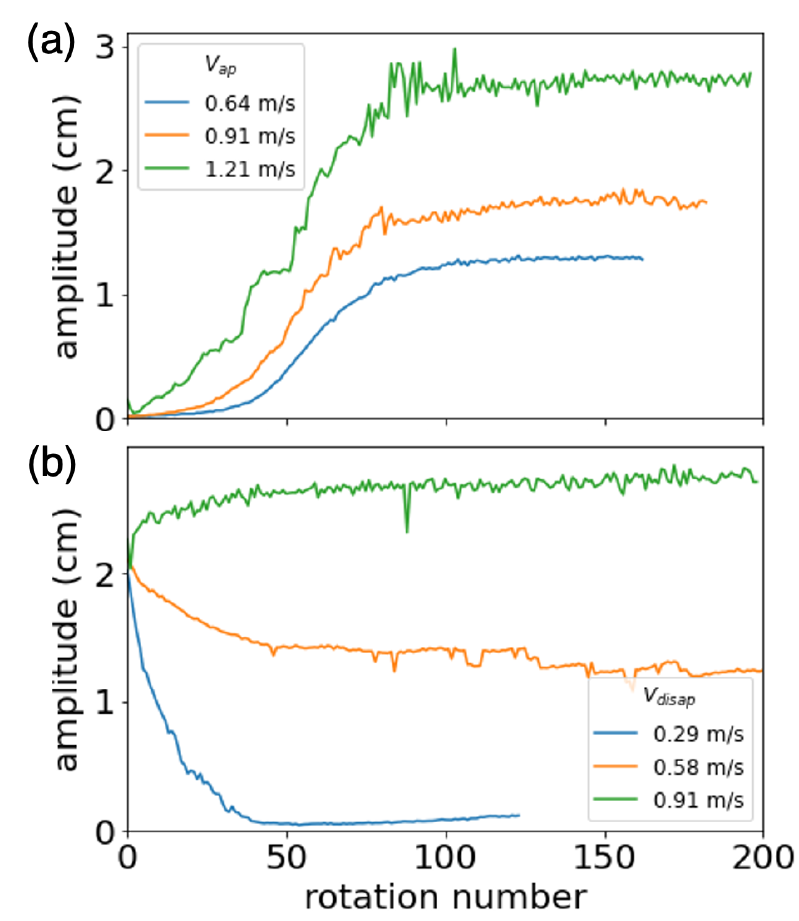}
  \caption{The variation of the amplitude as a function of rotation number. (a)~Various $v_\mathrm{ap}$(=0.64, 0.91, 0.21~m~s$^{-1}$) cases are presented. (b)~Fixed $v_\mathrm{ap}=1.21$~m~s$^{-1}$ and various $v_\mathrm{red}$(=0.29, 0.58, 0.91)~m~s$^{-1}$ cases are presented. The steadiness of the amplitude can be confirmed in the late stage.}
  \label{fig:ampwithrot}
\end{figure}

In this experiment, we investigate the steady ripple patterns. To check the steadiness of the ripple structure, we monitor the temporal development of the amplitude of ripple patterns. Figure~\ref{fig:ampwithrot} illustrates the example data of amplitude as a function of rotation number. Figure~\ref{fig:ampwithrot}(a) shows the amplitude of various $v_\mathrm{ap}$ values (in preparation stage). Figure~\ref{fig:ampwithrot}(b) shows the amplitude (in reducing stage) data with various $v_\mathrm{red}$ and fixed $v_\mathrm{ap}=1.21$~m~s$^{-1}$. All the amplitude data approach the asymptotic steady values as shown in Figure~\ref{fig:ampwithrot}. This tendency is confirmed in all experimental results. The amplitude and wavenumber analyzed in this study are measured in the steady regime. Specifically, the amplitude value is defined by the average of the maximum height difference in each wave. The data plotted in Figures~\ref{fig:resap},\ref{fig:amphys}, and \ref{fig:ampnorm} are the average amplitude of the final rotation of the plow.

\end{document}